\documentclass[10pt,superscriptaddress,pra,twocolumn]{revtex4}%
\usepackage{amsfonts}
\usepackage{amsmath}
\usepackage{amssymb}
\usepackage{graphicx}
\usepackage{graphics}
\usepackage[usenames]{color}%

\begin{document}
\title{Universal three-body recombination via resonant $d$-wave interactions}
\author{Jia Wang}
\affiliation{Department of Physics and JILA, University of Colorado, Boulder, CO 80309, USA}
\affiliation{Department of Physics, University of Connecticut, Storrs, CT 06269, USA}
\author{J. P. D'Incao}
\affiliation{Department of Physics and JILA, University of Colorado, Boulder, CO 80309, USA}
\author{Yujun Wang}\thanks{Present address: Joint Quantum Institute, University of Maryland and NIST, College Park, Maryland, 20742}
\affiliation{Department of Physics and JILA, University of Colorado, Boulder, CO 80309, USA}
\author{Chris H. Greene}
\affiliation{Department of Physics and JILA, University of Colorado, Boulder, CO 80309, USA}
\affiliation{Department of Physics, Purdue University, West Lafayette, IN 47907, USA}
\begin{abstract}
For a system of three identical bosons interacting via short-range forces, when two of the atoms are about to form a two-body $s$-wave dimer, the Efimov effect takes place leading to the formation of an infinite number of three-body (Efimov) states. The lowest Efimov state crosses the three-body break-up threshold when the $s$-wave two-body scattering length is $a \approx -9.73 r_{\rm vdW}$, $r_{\rm vdW}$ being the van der Waals length. This article focuses on a generalized version of this Efimov scenario, where two of the atoms are about to form a two-body $d$-wave dimer, resulting in strong $d$-wave interactions. Bo Gao has argued, in [Phys. Rev. A. {\bf 62}, 050702(R) (2000)], that for broad resonances the $d$-wave dimer is always formed when $a \approx 0.956 r_{\rm vdW}$. Our results demonstrate that a single universal three-body state associated with the $d$-wave dimer is also formed near the three-body break-up threshold at $a \approx 1.09 r_{\rm vdW}$, or alternatively $a_{2}=0.902r_{\rm vdW}$, where $a_{2}$ is the two-body $d$-wave scattering length. Such a universal three-body state is signaled experimentally by an enhancement of the three-body recombination rate. The three-body effective potential curves that are crucial for understanding the recombination dynamics are also calculated and analyzed. An improved method to calculate the couplings, effective potential curves, and recombination rate coefficients is presented.
\end{abstract}
\maketitle

\section{Introduction}
In recent years, the investigation of Efimov physics \cite{Efimov1971} has attracted much interest and has sparked a tremendous advance in our understanding of its fundamental aspects from both experimental \cite{Kraemer2006,Berninger2011,7Li_Rice,7Li_Kayk_A,7Li_Kayk_B,6Li_Selim_A,6Li_Selim_B,6Li_PenState_A,6Li_PenState_B,85Rb_JILA,RbK_LENS} and theoretical \cite{PetrovNarrowRes,yujun2011narrow,BraatenReview,Hammer2007,StecherNaturePhysics2009,Dipoles} viewpoints. Efimov physics describes a universal phenomenon in three-body systems: when the atoms are about to form a two-body $s$-wave dimer, i.e., the $s$-wave scattering length $a$ goes to infinity, and an infinite number of three-body Efimov states exists. The ratio between the energies of nearby Efimov states is given by a universal scaling factor $E_{n + 1} /E_n  = e^{-2\pi /s_0 } $ , where $s_0$ depends only on the mass ratios, number of resonant interactions and quantum statistics of the atoms, e.g.,  $e^{\pi /s_0 }  \approx 22.7$ for three identical bosons.

Therefore, a single parameter, called the three-body parameter, is needed to determine the whole Efimov spectrum and other scattering observables.  One possible definition of the three-body parameter is $a_{-}^{*}$, the value of $a$ at which the first resonance in the three-body recombination rate $K_3$ appears, i.e., at which the lowest Efimov state energy crosses the three-body break-up threshold ($E=0$). One fundamental assumption underlying all Efimov physics is that the three-body parameter encapsulates the short-range details of two- and three-body interactions, and thus should not be universal. Surprisingly, however, ultracold experiments with alkali atoms near Fano-Feshbach resonances \cite{Kraemer2006,Berninger2011,7Li_Rice,7Li_Kayk_A,7Li_Kayk_B,6Li_Selim_A,6Li_Selim_B,6Li_PenState_A,6Li_PenState_B,85Rb_JILA,RbK_LENS} have observed a universal value, namely $a_{-}^{*} \approx -9.1 r_{\rm vdW}$ in homonuclear atomic systems, where $r_{\rm vdW}$ is the van der Waals length. The van der Waals length $r_{\rm vdW}=\left({2 \mu_{2b} C_6} \right)^{1/4}/2$ is a characteristic length scale for the van der Waals interaction $-C_6/r^6$ between two neutral atoms with two-body reduced mass $\mu_{2b}$. A relevant energy scale, called the van der Waals energy, can be defined as $E_{\rm vdW}=\hbar^2/\left({2\mu_{2b} r_{\rm vdW}^2}\right)$.

This newly found universality in the three-body parameter was subsequently studied in several different theoretical models \cite{Schmidt,Chin3BP,Zinner3BP,Jia2012PRL,Ueda3BP}. In particular, Refs. \cite{Jia2012PRL,Ueda3BP} conclude that the universality of the three-body parameter comes from a universal barrier in the three-body potential curves arising from interatomic distances around 2 $r_{\rm vdW}$. This barrier originates from the suppression of the probability to find two atoms at distances shorter than $r_{\rm vdW}$ due to the increase of the classical local velocity and it is a universal property of van der Waals interactions between neutral atoms. Our previous studies \cite{Jia2012PRL,Yujun3BP} have indicated that the Lennard-Jones potential is an excellent model potential to study the universality of three-body physics in ultracold atomic gases, and for this reason it is also adopted in this analysis.

The present study, however, employs the Lennard-Jones potential model to study three-body systems with atom pairs close to a $d$-wave resonance. Interestingly, Gao \cite{Gao2000} predicted that $d$-wave dimers form at a universal value of $a$, $a \approx 0.956 r_{\rm vdW}$, for potentials with a van der Waals tail. Evidently, since that prediction was obtained for a single channel potential model its validity is expected to hold only for broad Fano-Feshbach resonances, and possibly is limited only to single-channel two-body interactions. Nevertheless, one natural question is: will there also be a universality of three-body physics near that value of scattering length, i.e., when $d$-wave interactions are expected to be strong? Here, the three-body recombination rate is studied to answer this question, and it is seen to exhibit two universal enhancements. Analysis of the three-body effective potential curves yields an intuitive understanding of these enhancements. One of these enhancements is caused by the simple fact that when the two-body $d$-wave state becomes barely bound an additional decay channel is formed, leading to an enhancement in recombination to $d$-wave dimers. The second enhancement, however, has a different nature. It corresponds to the formation of a universal three-body state that crosses the collision threshold for three free particles. Our analysis indicates that other three-body states associated with excited 2-body angular momenta are also possible, however, their signatures in recombination will be suppressed in the low energy limit due to the Wigner threshold laws for recombination \cite{Esry2001}. It is important to mention that since in experiments using ultracold atoms $s$- and $d$-wave states are coupled by the magnetic anisotropic dipole interaction the $s$-wave scattering actually diverges when the $d$-wave dimer becomes bound. Although the connection between these three-body states associated with $d$-wave resonant interactions and the multichannel physics in ultracold atoms still remains to be understood more deeply, the present study can offer an alternative parameterization in terms of the properties of the $d$-wave scattering length. Moreover, since the three-body effective potential curves are an important tool for understanding universality in few-body systems at ultracold temperatures~\cite{Jia2012PRL,Ueda3BP,Yujun3BP}, an improved method to calculate the coupling matrix elements is also developed in this paper.

The remainder of the paper is organized as follows. Section II tests the Gao prediction numerically for a Lennard-Jones potential, and demonstrates very good agreement. The three-body recombination rate is then calculated using the Lennard-Jones two-body potential with 2 $s$-wave bound states in the vicinity of the $d$-wave resonance. The improved numerical method is discussed in the subsection II.A. Analysis of the recombination rate using effective potential curves is made in subsection II.B. Section III checks the universality using the Lennard-Jones potential with 3 $s$-wave dimer bound states.

\section{Universal two-body $d$-wave dimer}
One of the key experimental tools used to study ultracold atomic gases is the Fano-Feshbach resonance, which can be utilized to magnetically tune the scattering length. Although the multichannel character of the hyperfine interactions leads to a great deal of complication, the single-channel van der Waals interactions has been shown to offer a good model for studying a broad Fano-Feshbach resonance in ultracold atomic gases \cite{ChinReview}. In 2000, Bo Gao predicted that for single-channel interactions with a van der Waals tail, $-C_6/r^6$, there is always a $d$-wave dimer (and dimers with higher angular momentum $l=4j+2$, where $j=1,2,3...$) that becomes bound at a universal value of the two-body $s$-wave scattering length $a=a^{*}=4\pi/[\Gamma(1/4)]^2 r_{\rm{vdW}} \approx 0.956 r_{\rm{vdW}}$ \cite{Gao2000}. [Note that $a^{*}$ has the same value as the so-called mean scattering length $\bar a$ as defined by Gribakin and Flambaum \cite{Flambaum1993,ChinReview}, which in turn should not be confused with the usual scattering length $a$.] Another special property of the two-body system at $a=a^{*}$ is that the effective range $r_{\rm{eff}}$ has the value $2 \left[ {\Gamma \left( {1/4} \right)} \right]^2/\left( {3\pi } \right) r_{\rm{vdW}} \approx 2.789 r_{\rm{vdW}}$, which equals the effective range when $s$-wave scattering length is infinity. In Gao's work \cite{Gao2000}, the universality of $a^{*}$ is based on an $l$-independent quantum-defect theory, where the short-range physics can be described by a single short-range parameter, $K^c$, which is itself approximately $l$-independent under proper conditions that are satisfied by most diatomic systems. In combination with solutions for the long-range interactions, $K^c$ uniquely determines both the scattering and the bound state properties of diatomic systems.

\begin{figure}[htbp]
\includegraphics[width=0.5 \textwidth]{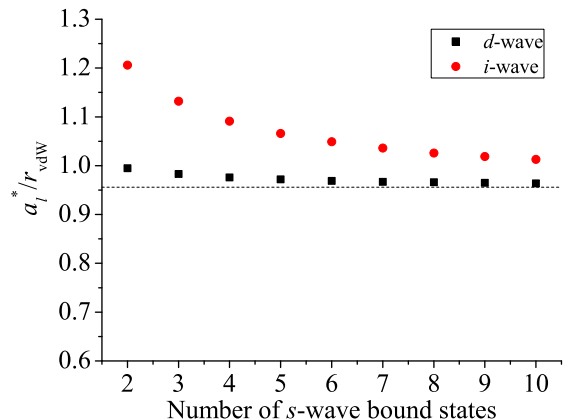}
\caption{(Color online) The values of the two-body $s$-wave scattering length $a^*_l$ at the point where a $d$-wave ($l=2$) dimer (black curve with square symbols) and where an $i$-wave ($l=6$) dimer just becomes bound (red curve with circular symbols) are shown as functions of the number of two-body $s$-wave bound states.}\label{astar}
\end{figure}

In this section, the universal value of the two-body $s$-wave scattering length $a_l^{*}$ [where a $d$-wave ($l=2$) or $i$-wave ($l=6$) dimer becomes bound] is studied numerically as a function of the number of two-body $s$-wave bound states and using the Lennard-Jones potential model for the two-body interaction:
\begin{equation}
v(r)=-\frac{C_6}{r^6}\left(1-\frac{\lambda^6}{r^6}\right).
\end{equation}
Note that in the present work the parameter $\lambda$ is used to adjust the values of $a$ as well as the number of bound states. (Some additional details about the Lennard-Jones potential are discussed in Appendix \ref{LJPotential}.) In Fig. \ref{astar}, the black square symbols (red circular symbols) indicates the numerical values of the two-body $s$-wave scattering length $a^*_2$ ($a^*_6$) at the point where a $d$-wave ($i$-wave) dimer becomes bound. The more $s$-wave bound states supported by the system, the deeper is the potential, implying different short-range physics. The universal prediction from Gao's work, $a^* \approx 0.956 r_{\rm vdW}$ \cite{Gao2000} (horizontal dashed line in Fig.~\ref{astar}), however, suggests that the actual form of the potential at short-distances is not important. In fact, our numerical results obtained for $a^*_l$ agree well with the Gao prediction of two-body binding energies [within $1\%$ ($6\%$) in the case of 10 $s$-wave bound states for a $d$-wave ($i$-wave)] and this agreement tends to improve as one increases the number of $s$-wave states. In the vicinity of the point where the dimer is just barely bound, i.e., when $a \le a^*_l$, the binding energy can be expressed as a linear function of the scattering length , i.e.,
\begin{equation}
E_l  / \left[{\hbar^2/(2 \mu_{2b} {r_{\rm{vdW}}^2})}\right]\approx  d_l \left( {a^*_l - a } \right)/r_{\rm{vdW}},
\end{equation}
where $d_2 \approx 5.8$ and $d_4 \approx 43$, for $d$-wave and $i$-wave states, respectively, are approximately universal. It is also interesting to note that near $a=a^*_2$ one expects strong $d$-wave interactions. In order to parametrize $d$-wave interactions in the ultracold regime, scattering properties must also be explored for two atoms colliding in the $l=2$ channel. In this case, due to the $1/r^6$ dependence of the Lennard-Jones potential, the $d$-wave elastic scattering phase-shift $\delta _2 \left( k \right)$ is known to have the following low energy expansion:
\begin{equation}
\tan \delta _2 \left( k \right) =  - \lambda_2 k^4  - a_2^5 k^5 ,
\end{equation}
\noindent
instead of the usual $k^{2l+1}$ dependence found for $l<2$. Here, $\lambda _2  = \left[ {\pi \Gamma \left( 5 \right)\Gamma \left( {1/2} \right)} \right]/\left[ {4\Gamma ^2 \left( 3 \right)\Gamma \left( {11/2} \right)} \right]r_{\rm vdW}^4 \approx 0.160 r_{\rm vdW}^4$ is a constant that only depends on the van der Walls length, and $a_2$ diverges when a $d$-wave state is just about to be bound \cite{Gianturco2006PRA}. Therefore, $a_2$ is denoted the ``$d$-wave scattering length'' in this study (which should not be confused with $a_2^*$), and is also used  in order to characterize our findings in terms of the $d$-wave interactions. This parameterization, therefore, allows a better comparison to experiments in ultracold quantum gases. It is important, however, to emphasize that the complicated multichannel structure of alkali atoms used in ultracold quantum gases can strongly affect the relations explored above between the three-body states found for $d$-wave interactions and the single channel parameters that characterize the present study. This is particularly true near narrow Feshbach resonances \cite{yujun2011narrow}.

\section{Three-body recombination}
This section focuses on the three-body recombination rate near the region where the $d$-wave dimer is about to be bound. In particular, the example of a Lennard-Jones potential with two $s$-wave bound states is elaborated. In this numerical example, the $d$-wave dimer becomes bound at about $a=0.995 r_{\rm{vdW}}$ and the $i$-wave dimer becomes bound at about $1.206 r_{\rm{vdW}}$. Later, the three-body recombination rate with more two-body $s$-wave bound states will be computed to test the universality of those results.

\subsection{Hyperspherical approach}
The hyperspherical approach is utilized here to study the three-body systems. The version of that approach adopted here has been discussed in detail in other recent studies \cite{JiaSVD,Jia2012PRL}. To briefly summarize, the three-body system is described in hyperspherical coordinates: a hyperradius $R$ that describes the overall size of the three-body system, two hyperangles $\theta$ and $\phi$ that describe the shape of the three-body system, and the three usual Euler angles $\alpha, \beta, \gamma$ that describe rigid-body rotations. In these coordinates, the three-body Schr\"odinger equation is written as,
\begin{equation}\label{SchrodingEq}
\left[ { - \frac{{\hbar ^2 }}{{2\mu _{3b} }}\left( {\frac{{\partial ^2 }}{{\partial R^2 }} - \frac{{\Lambda ^2 + 15/4}}{{R^2 }}} \right) + V\left( {R,\theta ,\varphi } \right)-E} \right]\psi _E  = 0,
\end{equation}
where $\Lambda^2\equiv\Lambda^2(\Omega)$ [$\Omega\equiv\{\theta,\varphi,\alpha,\beta,\gamma\}$] is the ``grand angular-momentum operator'' and $\mu_{3b}=m/\sqrt{3}$ is the three-body reduced mass of three identical atoms with mass $m$ \cite{Kendrick1999,Suno2002}.

\begin{figure}[htbp]
\includegraphics[width=0.5 \textwidth]{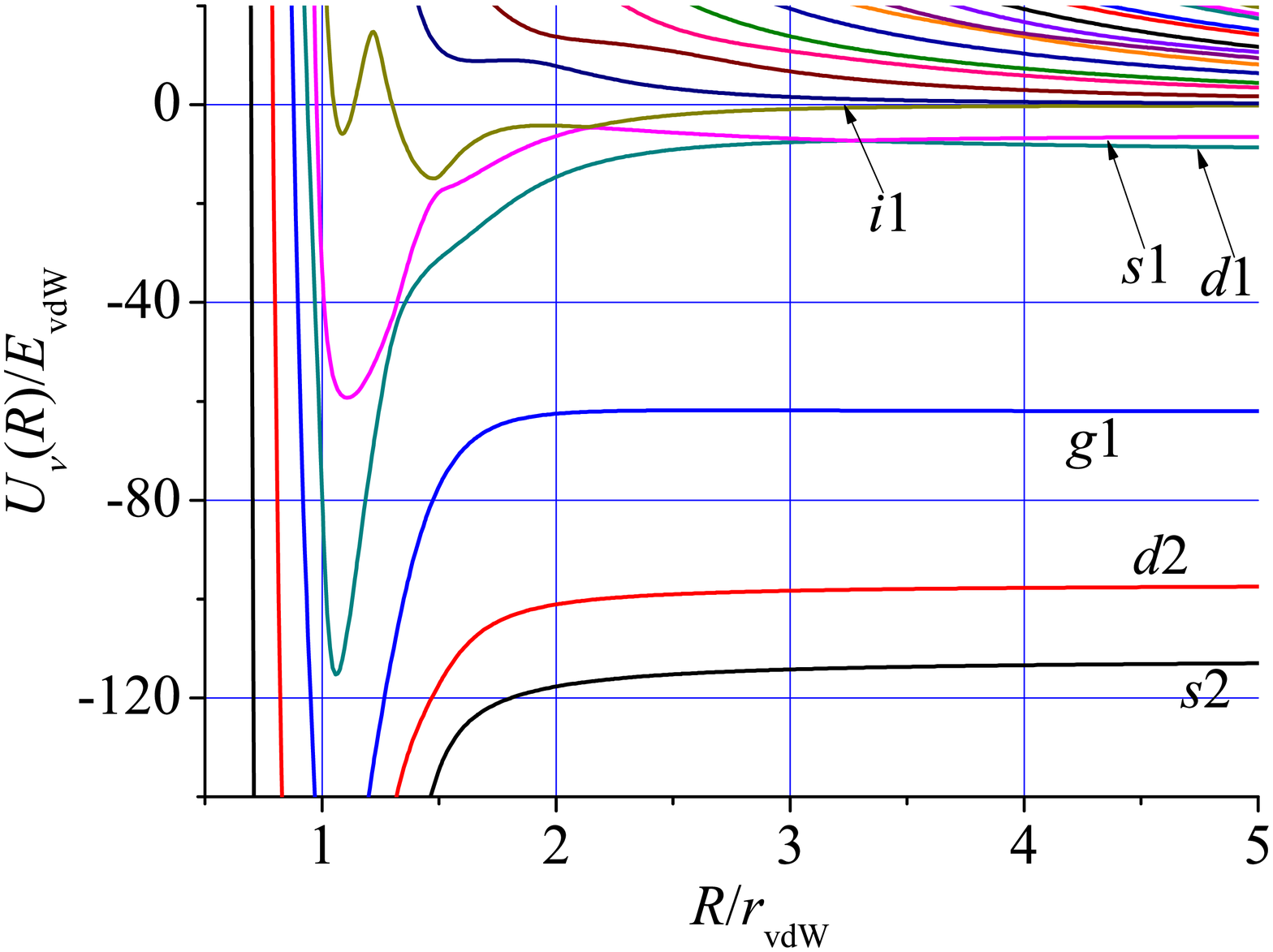}
\caption{(Color online) Adiabatic three-body potential curves for $a \approx 0.977 r_{\rm vdW}$. The $2+1$ channels, which corresponding to a dimer plus a free atom at very large distance, are labeled by a combination of a letter and a number. The letter denotes the angular momentum quantum number $l$ of the dimer, and the number labels the channels for the same dimer angular momentum from low-to-high dimer binding energies.}\label{Uad}
\end{figure}

To solve this Schr\"odinger equation, the first step in our method is to diagonalize the adiabatic Hamiltonian $H_{\rm{ad}}\left({R, \Omega}\right)$ and find the adiabatic potentials $U_{\nu}(R)$ and channel functions $\Phi _\nu  \left( {R;\Omega } \right)$:
\begin{equation}\label{AdiabaticEq}
H_{\rm{ad}}\left({R, \Omega}\right) \Phi _\nu  \left( {R;\Omega } \right) = U_\nu  \left( R \right)\Phi _\nu  \left( {R;\Omega } \right),
\end{equation}
whose solutions depend parametrically on $R$. The adiabatic Hamiltonian, containing {\em all} angular dependence and interactions, is defined as
\begin{equation}
H_{\rm{ad}}\left({R, \Omega}\right)=\left[ {\frac{{\hbar^2 \Lambda ^2 }}{{2\mu_{3b} R^2 }} + \frac{{15 \hbar^2}}{{8\mu_{3b} R^2 }} + V\left( {R,\theta ,\varphi } \right)} \right].
\end{equation}
For illustration, Fig. \ref{Uad} shows the three-body adiabatic hyperspherical potential curves for the case of $a \approx 0.977 r_{\rm{vdW}}$. [Note that in this figure the labels for the final recombination channels are also indicated and are useful for further analysis.]

The adiabatic potentials $U_{\nu}(R)$ and channel functions $\Phi_{\nu}(R;\Omega)$ obtained by solving Eq.~(\ref{AdiabaticEq}) for fixed values of $R$ contain all the correlations relevant to this problem. For each $R$, the set of $\Phi _\nu \left( {\Omega ;R} \right)$ is orthogonal,
\begin{equation}
\int {d\Omega \Phi _\mu  \left( {R;\Omega } \right)^* } \Phi _\nu  \left( {R;\Omega } \right) = \delta _{\mu \nu },
\end{equation}
and complete
\begin{equation}
\sum\limits_\tau  {\Phi _\tau  \left( {R;\Omega } \right)\Phi _\tau  \left( {R;\Omega '} \right)^*  = \delta \left( {\Omega  - \Omega '} \right)}.
\end{equation}
In practice, calculation of all the channel functions is time consuming and impractical. However, numerical studies have shown that only a small number of channels are needed as a truncated basis-set to expand the total wave function, e.g.,
\begin{equation}\label{ExpandAdiabatic}
\psi_E \left( {R,\Omega } \right) = \sum\limits_{\nu=1}^{N_c}  {F_\nu^E  \left( R \right)\Phi _{\nu} \left( {\Omega ;R} \right)},
\end{equation}
where $N_c$ is the number of channels adopted. Insertion of Eq. (\ref{ExpandAdiabatic}) into  Eq. (\ref{SchrodingEq}) leads to a set of coupled one-dimensional equations:
\begin{widetext}
\begin{equation}\label{RadialEq}
\left[ { - \frac{\hbar^2}{{2\mu_{3b}}}\frac{d^2}{{dR^2 }} + U_{\nu} \left( R \right)} -E \right]F_{\nu \nu '}^{E} \left( R \right) - \frac{\hbar^2}{{2\mu_{3b}}}\sum\limits_\mu  {\left[ {2P_{\nu \mu } \left( R \right)\frac{d}{{dR}} + Q_{\nu \mu } \left( R \right)} \right]} F_{\mu \nu '}^{E}\left( R \right)  = 0,
\end{equation}
\end{widetext}
where $\underline P$ and $\underline Q$ are the coupling matrices defined below, and $\nu '$ denotes the $\nu'-$th independent solution. (Hereafter, unless otherwise specified, we use an underline to denote the matrix form, e.g., $\underline P$ denotes a matrix with matrix element $P_{\nu\mu}$.) In the equation above, the non-adiabatic coupling matrices are defined as
\begin{equation}\label{Pmatrix}
P_{\nu \mu } \left( R \right) = \int {d \Omega } \Phi _\nu  \left( {R;\Omega } \right)^* \frac{\partial}{{\partial R}}\Phi _\mu  \left( {R;\Omega } \right),
\end{equation}
\begin{equation}
Q_{\nu \mu } \left( R \right) = \int {d\Omega } \Phi _\nu  \left( {R;\Omega } \right)^* \frac{{\partial^2 }}{{\partial R^2 }}\Phi _\mu  \left( {R;\Omega } \right),
\end{equation}
\noindent
and are responsible for inelastic scattering processes as well as the finite width of three-body resonant states. In practice, only the
\begin{equation}\label{P2matrix}
P_{\nu \mu }^2 \left( R \right) = - \int {d \Omega } \frac{\partial}{{\partial R}}\Phi _\nu  \left( {R;\Omega } \right)^* \frac{\partial}{{\partial R}}\Phi _\mu  \left( {R;\Omega } \right),
\end{equation}
component of $Q_{\nu \mu }$ is needed to solve the coupled equations \cite{JimBurkeThesis}. The relation between $\underline Q$ and $\underline P^2$ is given by $\frac{d}{{dR}}\underline P  =  - \underline {P^2 }  + \underline Q$. From the definition of the $\underline P$ and $\underline Q$ matrices, it is easy to see that the coupling matrices have the following properties: $P_{\nu \mu } = -P_{\mu \nu }$ and $P_{\nu \mu }^2 = P_{\mu \nu }^2$, which leads to $P_{\nu \nu} = 0$, and $Q_{\nu \nu } = P_{\nu \nu }^2$. In the hyperspherical adiabatic representation, the effective potentials are usually defined as
\begin{equation}
\widetilde U_\nu  \left( R \right) \equiv W_{\nu \nu } \left( R \right) = U_\nu  \left( R \right) - \frac{{\hbar^2 }}{{2\mu _{3b} }}P_{\nu \nu }^2 \left( R \right),
\end{equation}
and differs for the hyperspherical potential $U_{\nu}(R)$ by the addition of the diagonal correction $P_{\nu\nu}^2(R)$. The effective potentials are usually more physical than the raw Born-Oppenheimer potentials because they include hyperadial kinetic energy contributions through the $P_{\nu\nu}^2$ term. For example, the effective potential gives physical asymptotic behaviors of the system at large $R$ with the correct coefficient of $R^{-2}$. In some cases, one can even neglect the nonadiabatic couplings between different adiabatic channels and approximately solve the Schr\"{o}dinger equation as a single channel problem:
\begin{equation}\label{singlechannelapproximation}
\left[ { - \frac{{\hbar ^2 }}{{2\mu _{3b} }}\frac{{d^2 }}{{dR^2 }} + \tilde U_\nu  \left( R \right) - E} \right]F_{\nu \nu }^E \left( R \right) = 0,
\end{equation}
called the adiabatic approximation \cite{Kokoouline2003}. Although a semi-quantitative picture can emerge within this approximation, the fully coupled system of equation has to be solved to obtain a more quantitative description of the properties of the three-body system including inelasticity in particular.

Quantitative solutions of Eq.~(\ref{RadialEq}) hinge on the ability to calculate non-adiabatic couplings accurately. A traditional method for calculating the coupling matrices is to apply a simple finite difference scheme for the derivative of $\Phi _\mu  \left( {R;\Omega } \right)$, i.e.,
\begin{equation}\label{FiniteDiff}
\frac{\partial }{{\partial R}}\Phi _\mu  \left( {R;\Omega } \right) \approx \frac{{\Phi _\mu  \left( {R + \Delta R;\Omega } \right) - \Phi _\mu  \left( {R- \Delta R;\Omega } \right)}}{{2\Delta R}}.
\end{equation}
This scheme, however, is only accurate up to the first order in $\Delta R$. Here, we use instead an improved method to calculate the $\partial \Phi _\mu  \left( {R;\Omega } \right)/\partial R$. This method only relies on $H_{\rm{ad}}\left({R, \Omega}\right)$ and $\partial H_{\rm{ad}}\left({R, \Omega}\right) / \partial R$ at the desired value of $R$ and is shown to be in principle numerically exact (see Appendix \ref{AppendixPQ} for a detailed discussion).

\subsection{Three-body recombination rate}
\begin{figure}[htbp]
\includegraphics[width=0.5 \textwidth]{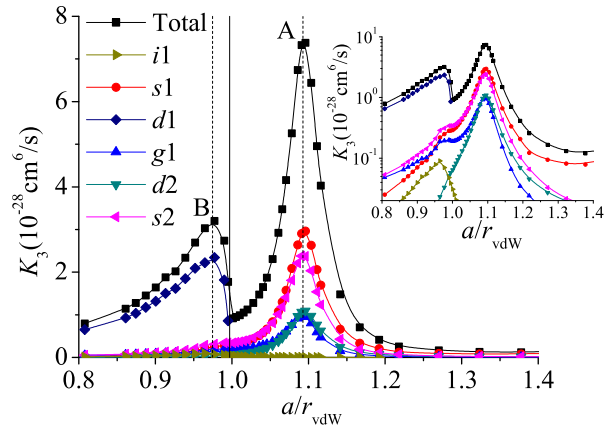}
\caption{(Color online) Total and partial $J^\pi=0^+$ three-body recombination rates as functions of the two-body scattering length $a$, shown on a linear scale, near a $d$-wave resonance associated with the $d$-wave dimer formation. The inset shows the same graph with a logarithmic scale for the y-axis. The units of $K_3$ are converted to $\rm{cm^6/s}$ by using the van der Waals length $r_{\rm vdW}= 101.0$ bohr and mass $132.905429$ amu of $^{133}$Cs. The solid vertical line indicates the value of the $s$-wave scattering length at the point where the $d$-wave dimer becomes bound. The two dashed vertical lines indicate the two values of $a$ at which recombination rate is enhanced, denoted A and B respectively.}\label{K3dwave}
\end{figure}

The numerical method described in the previous subsection yields more accurate coupling matrix elements and hence improved effective potential curves. These effective potentials are not only crucial for intuitively understanding the system, but also for quantitative calculations of three-body observables like the recombination rate. Together with the $R$-matrix propagation method discussed in Ref. \cite{JiaSVD}, the three-body recombination rate for $J^{\pi}=0^+$ symmetry, where $J$ is the three-body total angular momentum and $\pi$ the parity, are calculated and shown in Fig. \ref{K3dwave}. The solid black curve with square symbols shows the total recombination rate as a function of the scattering length $a$, near the $d$-wave resonance, i.e., for values of  $a$ near $a^{*}_2$. The total rate exhibits two clear enhancements labeled by ``A'' ($a_{\rm A} \approx 1.09 r_{\rm{vdW}}$) and ``B'' ($a_{\rm B} \approx 0.98 r_{\rm{vdW}}$). The enhancements ``A'' and ``B'', indicated in Fig.~\ref{K3dwave} by the vertical dashed lines, occur right after and right before the $d$-wave dimer becomes bound (see vertical solid line in Fig.~\ref{K3dwave}), respectively. Analysis of the partial recombination rates helps to elucidate the mechanism responsible for these enhancements. The enhancement around the peak ``A'' appears in all the partial three-body recombination rates. After the $d$-wave becomes bound and forms a new decay channel, the partial $K_3$ rates near the enhancement ``B'' have a different characteristic. The total rate is now dominated only by the partial rate into this new $d$-wave channel [see the inset of Fig. \ref{K3dwave}]. These features suggest that peak ``A'' might be a resonance due to a three-body state that crosses the three-body threshold, while peak ``B'' seems to relate to some other threshold behavior related to the formation of the new $d$-wave decay channel. The effective potentials near these enhancements, however, can yield insights into their origin.

\begin{figure}[htbp]
\includegraphics[width=0.5 \textwidth]{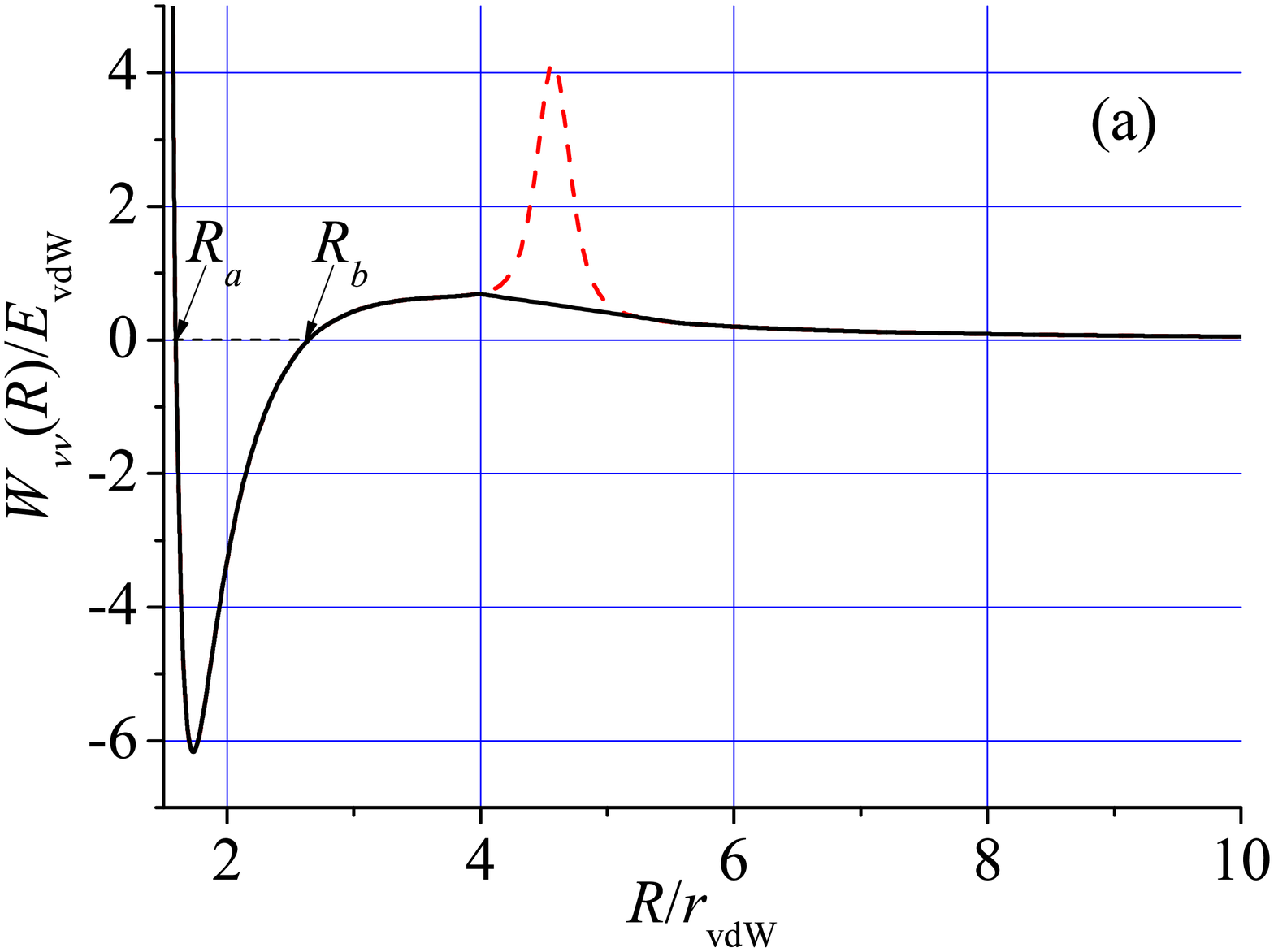}
\includegraphics[width=0.5 \textwidth]{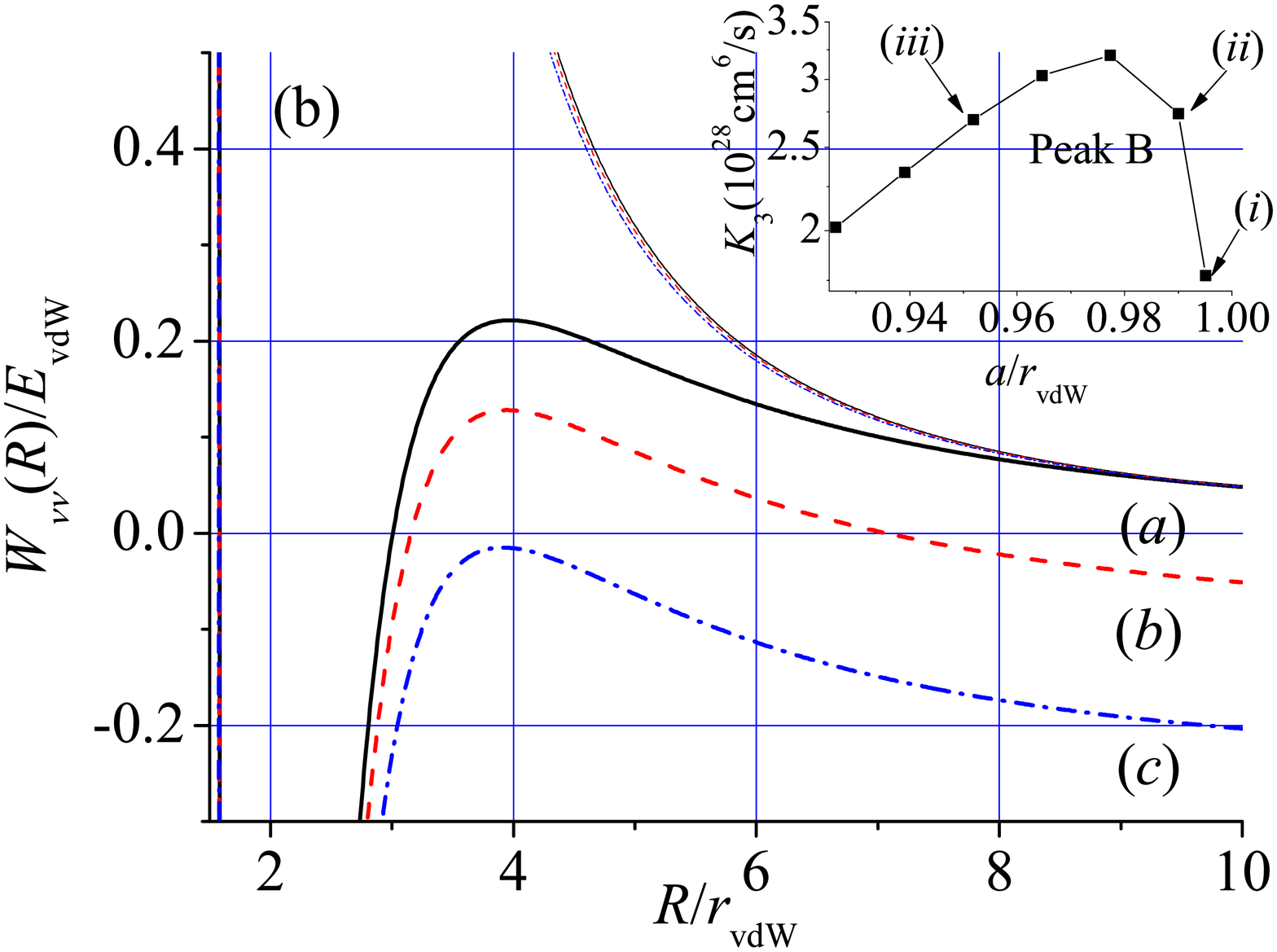}
\caption{(a) (Color online) $W_{\nu \nu}(R)$ as a function of hyperradius $R$ at the $K_3$ resonance peak A. (b) (Color online) $W(R)_{\nu \nu}$ as a function of hyperradius $R$ around the $K_3$ enhancement peak B. The inset shows the three points where the potential curves correspond to.}\label{WAB}
\end{figure}

First, consider the physical origin of the enhancement A. Figure. \ref{WAB} (a) shows the adiabatic hyperspherical potential curves at peak A. Only the channels relevant to the resonances are shown in this figure. The red dashed-line shows the effective hyperspherical potential. We have diabatized the potential near an avoided crossing since it plays no role, and display it as the black solid curve. The black solid curve shows an outer barrier, and a three-body short-ranged state could be supported for this potential. To check whether there is a three-body state near the threshold, a calculation of the WKB phase at zero scattering energy is carried out:
\begin{equation}
\phi _{\rm WKB}  = \int_{R_a }^{R_b } {\frac{\hbar }{{\sqrt { - 2\mu _{3b} W_{\nu \nu } \left( R \right)} }}} dR,
\end{equation}
where $R_a$ and $R_b$ are the classical turning points shown in Fig. \ref{WAB} (a). The calculated WKB phase $\phi _{\rm WKB}$ is about $0.51 \pi$, which is strong evidence for the existence of a three-body state. [Below, we calculate the actual energy of this state and describe some of its properties.] It is this three-body state associated with the $d$-wave dimer state crossing the three-body break-up threshold that causes the enhancement A in the three-body recombination rate. The physical picture is that the three-body state at threshold can trap the system within the potential well for a long time, hence making it more likely that the system will decay into deeper channels. Therefore, this decay process only depends on the initial channel, which is consistent with the fact noted above that near the enhancement A all partial rates display the resonance effect.

Next consider the enhancement mechanism responsible for peak B. Figure \ref{WAB} (b) shows the effective potentials $W_{\nu\nu}(R)$ as functions of $R$ for three different scattering lengths around the enhancement peak B as solid curves, dashed curves, and dash-dotted curves. The corresponding $K_3$ rates are indicated by the three points $(i)$, $(ii)$, and $(iii)$ in the inset of Fig. \ref{WAB}. Here, the dominant pathway at low energy is from the lowest entrance channel to the newly formed $d$-wave channel. For each scattering length, the thin curves in Fig.~\ref{WAB} are the entrance channels, and as one can readily see, they do not change substantially in all three cases. This means that the incident channel potential curve depends on the atom-atom scattering length only weakly. The thick curves in Fig.~\ref{WAB} are the most important recombination channel associated with the new $d$-wave channel. It is clear that from point $(i)$ to $(iii)$, the potential barrier decreases appreciably and can explain the increase of recombination as it gradually allows the three-body wave function to approach shorter distances, where non-adiabatic couplings are expected to be stronger.

From Ref. \cite{JiaSVD}, the scattering matrix element describing the recombination from the lowest incident channel to the newly formed $d$-wave channel can be approximated by,
\begin{equation}
S_{fi}  \approx 2\pi i\int_0^\infty  {dRF_f \left( R \right)W_{fi} \left( R \right)F_i \left( R \right)}
\end{equation}
where $F_i$ ($F_f$) are the regular solutions of Eq. (\ref{singlechannelapproximation}) --- single channel approximation --- for the initial (final) channel $i$ ($f$). As mentioned above, the incident channel is dominated by the centrifugal barrier which, in turn, is independent of the scattering length. Hence the energy-normalized $F_i$ are well approximated by Bessel functions:
\begin{equation}
F_i \left( R \right) \approx \sqrt {\mu _{3b} R} J_2 \left( {k_i R} \right),
\end{equation}
which is, of course, also independent of scattering length. The coupling matrix element $W_{fi} \left( R \right)$ is also found, numerically, to depend on scattering length only weakly for the three cases we studied here. Therefore, we assume that the product of $W_{fi} \left( R \right)F_i \left( R \right)$ does not depend appreciably on the scattering length. In addition, this product is shown numerically to be significant only at small hyperradius (around 1.3 $r_{\rm vdW}$ to 2.8 $r_{\rm vdW}$). Therefore, the scattering length dependence of $S_{fi}$ is fully determined by $F_f\left( R \right)$ at short distances. There are two competing mechanisms controlling the amplitude of $F_f\left( R \right)$: one is the tunneling through the outer barrier, which makes the amplitude become increasingly larger, as the barrier becomes increasingly lower from $(i)$ to $(iii)$; the other is the overall amplitude for an energy normalized wave function, which makes the amplitude becomes increasingly smaller, since the threshold energy associated to the energy of the $d$-wave dimer become lower and lower, and the out-going wave vector become larger and larger. These two competing mechanisms help to form peak structure B. The WKB phase calculated for these three cases is smaller than $\pi/2$, which is another evidence that this enhancement is {\em not} due to a three-body state resonance. This is also consistent with the fact that such an enhancement is not shown in the partial recombination rates for channels other than the newly formed $d$-wave.

\section{Three-body state associated with the $d$-wave dimer}
\begin{figure}[htbp]
\includegraphics[width=0.5 \textwidth]{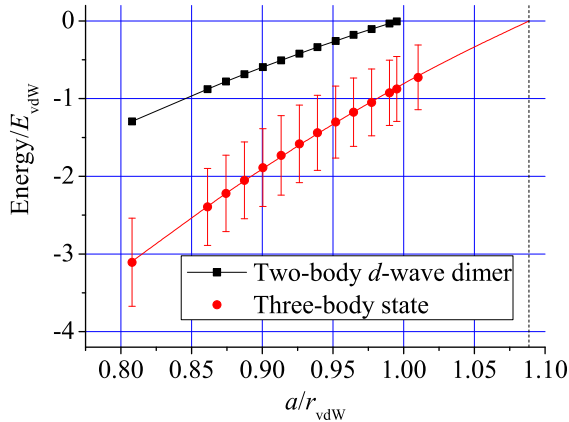}
\caption{(Color online) Energy of the three-body bound state associated with a $d$-wave dimer as a function of scattering length $a$, both in van der Waals units.}\label{bs3}
\end{figure}

While the Efimov states can be viewed as three-body states associated with a $s$-wave dimer near the three-body breakup threshold, the new three-body state discussed in the previous section is a three-body state that can be associated with a $d$-wave dimer. Because of the existence of deeper atom-dimer thresholds, this three-body state is actually a quasi-bound state. Figure \ref{bs3} shows the three-body state energy as a function of the scattering length $a$ (red dots), with the error bars indicating the width of the quasi-bound state. The red line represents a fitting formula to this energy:
\begin{equation}
\frac{E_{3b} }{ \hbar^2/\left({2\mu_{2b} r_{\rm{vdW}}^2}\right)}=  d_{3b} \frac{\left( {a^*_{3b} - a} \right)}{r_{\rm vdW}} + e_{3b} \frac{\left( {a^*_{3b} - a } \right)^2}{r_{\rm vdW}^2},
\end{equation}
where $a^*_{3b} \approx 1.09 r_{\rm{vdW}}$, $d_{3b} \approx 8.21$ and $e_{3b} \approx 10.01$ are fitting parameters. For comparison, the figure also shows the $d$-wave energy as the black curve with solid square symbols. The three-body energy crosses the three-body breakup threshold at $a \approx 1.09 r_{\rm{vdW}}$, which corresponds to peak A in $K_3$. As one can see, the three-body state is formed before the $d$-wave state becomes bound and its energy is always below the $d$-wave dimer energy (the black curve with square symbols in Fig. \ref{bs3}). The difference between the dimer and trimer energy also increases when the scattering length becomes smaller, and the dimer becomes deeper. However, the width of the three-body quasi-bound state does not seem to change appreciably.

The Gao analysis in Ref.~\cite{Gao2000}, suggests that there is also an $i$-wave dimer that becomes bound at a nearby scattering length. However, the $i$-wave dimer does not affect these two peaks. The $i$-wave channel has a very sharp avoided crossing with all the other channels. As a consequence, the partial $K_3$ rate into the $i$-wave dimer channel is negligible compared with the other partial rates. One explanation may be that forming a $i$-wave ($l=6$) dimer results in the exchange of a large amount of angular momentum between the dimer and the free atom. Therefore, although higher partial waves $l=10,14,18...$ might also be formed at nearby scattering lengths, they are not expected to show any strong features in the three-body recombination rates at ultracold energies.

\begin{figure}[htbp]
\includegraphics[width=0.5 \textwidth]{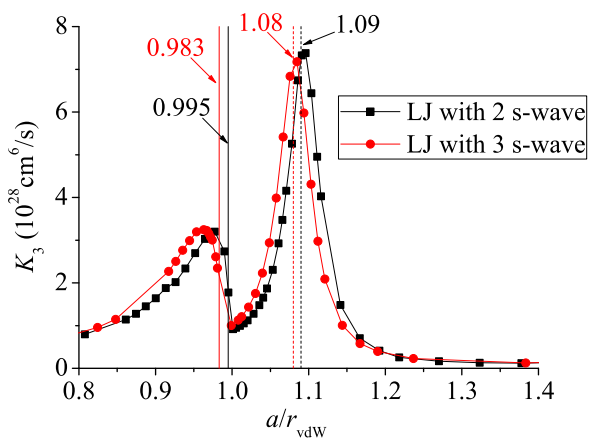}
\caption{(Color online) The enhancements for the total three-body recombination rates at about $a=0.995 r_{\rm vdW}$ for Lennard-Jones potential with 2 and 3 $s$-wave bound states. $K_3$ is convert to $\rm{cm^6/s}$ by using van der Waals length $r_{\rm vdW}= 101.0$ bohr and mass $132.905429 $ amu of $^{133}$Cs}\label{K3dwaveLJ23}
\end{figure}

\begin{figure}[htbp]
\includegraphics[width=0.5 \textwidth]{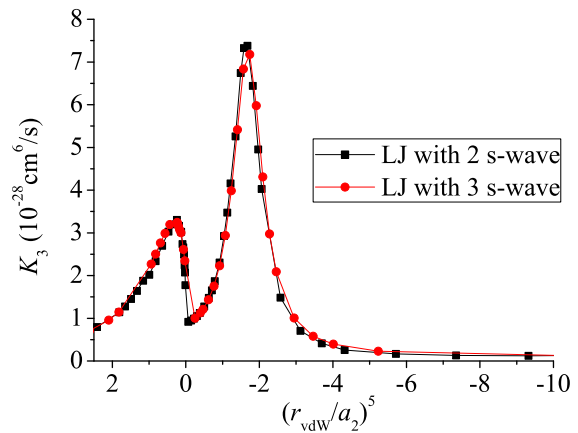}
\caption{(Color online) The total three-body recombination rates as a function of $r_{\rm vdW}^5/a_2^5$. The units of $K_3$ are the same as Fig. \ref{K3dwaveLJ23}.}\label{K3dwavea2}
\end{figure}

Finally, to test the universality of our results, the three-body recombination enhancements for the model potentials of Lennard-Jones type, having either 2 or 3 $s$-wave bound states, are shown in Fig. \ref{K3dwaveLJ23}. In this figure, the red (black) solid vertical line shows where the $d$-wave dimer crosses the threshold for LJ with 2 (3) $s$-waves. The red (black) dashed vertical line indicates where the three-body state associated with the $d$-wave level crosses the threshold for LJ with 2 (3) $s$-wave states. The values of the $s$-wave scattering length for such enhancements from these two different models differs by only 0.01 $r_{\rm vdW}$. This small difference is consistent with the small shift of the scattering length value where the $d$-wave dimer crosses the three-body threshold (see Fig.~\ref{astar}). As discussed before, the $d$-wave scattering length is a better parameter to describe $d$-wave interactions. We have therefore analyzed the three-body recombination rate as a function of $1/a_2^5$, which is shown in Fig. \ref{K3dwavea2}. The agreement between calculations using different potential models in fact looks better in this parameterization, where the three-body state crosses the threshold at $a_2 \approx 0.902 r_{\rm vdW}$ for both LJ with 2 and 3 $s$-waves bound states.

This result suggests (although it does not prove rigorously) that the three-body state associated with the $d$-wave dimer is also universal. We also note that at the threshold for formation of the three-body state, the effective potential relevant for that state does also display an inner repulsive barrier [see Fig. \ref{WAB} (a)] preventing atoms from approaching to hyperradii smaller than $1.5r_{\rm vdW}$. This indicates that the properties of such a three-body state should also be insensitive to the details of nonadditive short-range three-body interactions.

\section{Summary}
In summary, we have calculated the three-body recombination rates for Lennard-Jones potentials with 2 and 3 $s$-wave bound states for small positive values of the scattering length. Two universal enhancement peaks are found at about $a=0.995$ $r_{\rm vdW}$. In particular, one of the enhancement peaks corresponds to a new universal three-body state that is associated with a $d$-wave dimer and is expected to exist in general for potentials having  a van der Waals tail.

\section*{Acknowledgements}
The authors want to thank Eric A. Cornell, John Bohn, Cheng Chin, Robin C\^{o}t\'{e} and Doerte Blume for helpful discussions. This work was supported in part by NSF and the AFOSR-MURI.
\appendix
\section{Lennard-Jones potential}\label{LJPotential}
\begin{figure}[htbp]
\includegraphics[width=0.5 \textwidth]{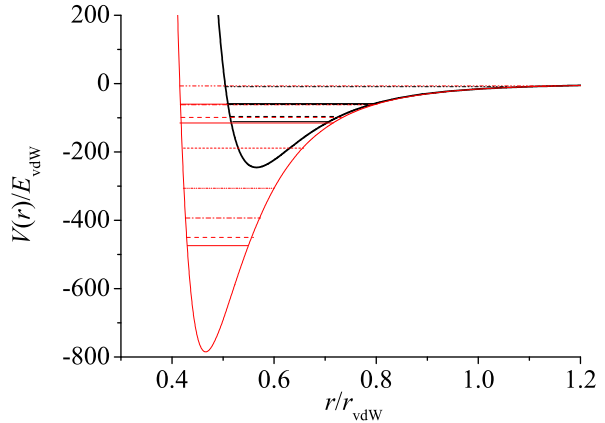}
\caption{(Color online) Lennard-Jones potential with $\lambda \approx 0.503 r_{\rm vdW}$ (black thicker curve) and $\lambda \approx 0.414 r_{\rm vdW}$ (red thinner curve). The horizontal lines show the bound state spectrum with even angular momentum quantum number $l$. The line style solid, dashed, dash-dotted, dash-dot-dotted, and short-dashed indicates $l=0,2,4,6,8$ correspondingly.}\label{LJPot}
\end{figure}

The two-body model potentials applied here are the Lennard-Jones potential, which can be expressed as a function of interatomic distances $r$:
\begin{equation}
V\left( r \right) =  - \frac{{C_6 }}{{r^6 }}\left( {1 - \frac{{\lambda ^6 }}{{r^6 }}} \right).
\end{equation}
This form of potential has a van der Waals tail $-C_6/r^6$ at long range and a repulsive core near the origin controlled by the parameter $\lambda$. In the present study, the values of $\lambda$ are adjusted to give the desired $s$-wave scattering length $a$ and number of bound states. Figure \ref{LJPot} shows two Lennard-Jones potentials with a zero-energy $d$-wave state and their corresponding spectrum of bound state with even $l$. (The zero energy $d$-wave state is not shown in Fig. \ref{LJPot}). The black thicker curve corresponds to $\lambda \approx 0.503 r_{\rm vdW}$ and has two $s$-wave bound states with $a \approx 0.995 r_{\rm vdW}$. The red thinner curve corresponds to $\lambda \approx 0.414 r_{\rm vdW}$ and has three $s$-wave bound states with $a \approx 0.983 r_{\rm vdW}$. It is shown that although the potentials at short-range are very different for the two different parameters, the scattering length and the spectrum near threshold are similar.

\section{Improved method for calculating nonadiabatic coupling matrices}\label{AppendixPQ}
Equation (\ref{FiniteDiff}) is only accurate up to the first order in $\Delta R$. In addition, the value chosen for $\Delta R$ in a realistic numerical calculation can sometimes be tricky.
When $\Phi _\mu  \left( {R;\Omega } \right)$ changes very rapidly, e.g., near a sharp avoided crossing, we need to choose
a very small step size $\Delta R$. In contrast, when $\Phi _\mu  \left( {R;\Omega } \right)$ changes very slowly, e.g., at
very large distances $R$, we need to choose a relatively larger step size $\Delta R$, or else
$\Phi _\mu  \left( {R + \Delta R;\Omega } \right) - \Phi _\mu  \left( {R- \Delta R;\Omega } \right)$ would be too small,
and the accuracy would be limited by the machine precision.

One way to improve the accuracy is to apply the Hellmann-Feynman theorem. This theorem can give us analytical formulas for the coupling matrices if we know the derivative of the adiabatic Hamiltonian $\frac{\partial }{{\partial R}}H_{\rm{ad}}$ from the following derivation. First, taking the derivative of both sides of Eq. (\ref{AdiabaticEq}) leads to
\begin{eqnarray}\label{DeAdiabaticEq}
&& \left[ {H_{\rm{ad}} \left( {R,\Omega } \right) - U_\nu  \left( R \right)} \right]\frac{\partial }{{\partial R}}\Phi _\nu  \left( {R;\Omega } \right) =  \nonumber \\ && - \left[ {\frac{\partial }{{\partial R}}H_{\rm{ad}} \left( {R,\Omega } \right) - \frac{\partial }{{\partial R}}U_\nu  \left( R \right)} \right]\Phi _\nu  \left( {R;\Omega } \right).
\end{eqnarray}
Next, multiplying $\Phi _\mu  \left( {R;\Omega } \right)^*$ on both sides of Eq. (\ref{DeAdiabaticEq}) and integrating over $\Omega$ gives
\begin{eqnarray}
P_{\mu \nu} &=& \int {d\Omega \Phi _\mu  \left( {R;\Omega } \right)^* } \frac{\partial }{{\partial R}}\Phi _\nu  \left( {R;\Omega } \right) \nonumber \\ &=& - \frac{\int {d\Omega \Phi _\mu  \left( {R;\Omega } \right)^* \left[{\frac{\partial }{{\partial R}}H_{\rm{ad}} \left( {R,\Omega } \right)}\right]\Phi _\nu  \left( {R;\Omega } \right)}}{\left[ {U_\mu  \left( R \right) - U_\nu  \left( R \right)} \right]},
\end{eqnarray}
where $ \mu \ne \nu$, and
\begin{equation}
\frac{\partial }{{\partial R}}U_\nu  \left( R \right) = \int {d\Omega \Phi _\nu  \left( {R;\Omega } \right)^* } \left[ {\frac{\partial }{{\partial R}}H_{\rm{ad}} \left( {R,\Omega } \right)} \right]\Phi _\nu  \left( {R;\Omega } \right)
\end{equation}
after some manipulation of algebra. The Hellmann--Feynman theorem is in principle exact. The matrix elements for $P^2$ can be obtained by
\begin{equation}
P_{\mu \nu }^2  = \sum\limits_{\tau  = 1}^{N_c } {P_{\mu \tau } P_{\tau \nu } },
\end{equation}
where $N_c$ is the number of channels. However, numerical studies show that the convergence of $P_{\mu \nu }^2$ with respect to number of channels is very slow, making this method impractical.

We now introduce an improved method to calculate $\frac{\partial }{{\partial R}}\Phi _\mu  \left( {R;\Omega }\right)$. It is in principle numerically exact. The first hint of the derivation of this method is that Eq. (\ref{DeAdiabaticEq}) seems plausible to be directly solved for $\frac{\partial }{{\partial R}}\Phi _\mu  \left( {R;\Omega } \right)$ by
\begin{eqnarray}
\frac{\partial }{{\partial R}}\Phi _\nu  \left( {R; \Omega } \right) =&&  - \left[ {H_{\rm{ad}} \left( {R,\Omega } \right) - U_\nu  \left( R \right)} \right]^{ - 1} \times \\ &&\left[ {\frac{\partial }{{\partial R}}H_{\rm{ad}} \left( {R,\Omega } \right) - \frac{\partial }{{\partial R}}U_\nu  \left( R \right)} \right]\Phi _\nu  \left( {R;\Omega } \right) \nonumber .
\end{eqnarray}
However, this solution is forbidden since ${H_{\rm{ad}} \left( {R,\Omega } \right) - U_\nu  \left( R \right)}$ is singular: $\left| {H_{\rm{ad}} \left( {R,\Omega } \right) - U_\nu  \left( R \right)} \right| = 0$, meaning that ${H_{\rm{ad}} \left( {R,\Omega } \right) - U_\nu  \left( R \right)}$ is not invertible. The singularity can also be understood from the fact that the equation
\begin{eqnarray}\label{EqDeAdiabatic}
&& \left[ {H_{ad} \left( {R,\Omega } \right) - U_\nu  \left( R \right)} \right]\chi _\nu  \left( {R;\Omega } \right) =  \\ &&- \left[ {\frac{\partial }{{\partial R}}H_{ad} \left( {R,\Omega } \right) - \frac{\partial }{{\partial R}}U_\nu  \left( R \right)} \right]\Phi _\nu  \left( {R;\Omega } \right) \nonumber
\end{eqnarray}
does not have a unique solution, $\chi _\nu  \left( {R; \Omega } \right)$. In fact, any functions with the form of
\begin{equation}\label{DeAdiabaticEqSolution}
\chi _\nu  \left( {R;\Omega } \right) = \frac{\partial }{{\partial R}}\Phi _\nu  \left( {R;\Omega } \right) + c \Phi _\nu  \left( {R;\Omega } \right),
\end{equation}
(where $c$ is an arbitrary number) can be a solution of Eq. (\ref{EqDeAdiabatic}). The singularity of matrix ${H_{\rm{ad}} \left( {R,\Omega } \right) - U_\nu  \left( R \right)}$ can be removed by considering the additional condition that
\begin{equation}
\int {d\Omega \Phi _\nu  \left( {R,\Omega } \right)^* } \frac{\partial }{{\partial R}}\Phi _\nu  \left( {R,\Omega } \right) = 0,
\end{equation}
which can be derived from the normalization condition of $\Phi _\nu  \left( {R,\Omega } \right)$, as shown in Ref. \cite{Nelsen1976}. Nevertheless, our numerical studies show that even without removing the singularity, the use of numerical solution packages such as ``Linear Algebra PACKage'' (LAPACK) \cite{LAPACK} or PARDISO \cite{PARDISO} to solve Eq. (\ref{EqDeAdiabatic}) directly still gives an accurate solution $\chi _\nu  \left( {R;\Omega } \right)$ in the form of Eq. (\ref{DeAdiabaticEqSolution}) with an unknown $c$. And once we have the numerical solution $\chi _\nu  \left( {R;\Omega } \right)$, $c$ can be calculated by
\begin{equation}
c=\int {d\Omega \Phi _\nu  \left( {R;\Omega } \right)^* } \chi _\nu  \left( {R;\Omega } \right).
\end{equation}
Finally, the derivative of $\Phi _\nu  \left( {R;\Omega } \right)$ can be written as
\begin{equation}
\frac{\partial }{{\partial R}}\Phi _\nu  \left( {R;\Omega } \right) = \chi _\nu  \left( {R;\Omega } \right) - c \Phi _\nu  \left( {R;\Omega } \right),
\end{equation}
which can be inserted into Eq. (\ref{Pmatrix}) and Eq. (\ref{P2matrix}) for the coupling matrices. The $P$ matrices obtained in this way are found to be numerically the same as the one calculated from the Hellmann--Feynman theorem up to machine precision, proving that our $\frac{\partial }{{\partial R}}\Phi _\nu  \left( {R;\Omega } \right)$ are numerically accurate. Therefore, this method can give numerically very accurate coupling matrices. The major limitation of this approach is that it becomes computationally more expensive to obtain nonadiabatic couplings as the number of requested channels increases. Nevertheless, for our typical calculations the number of relevant three-body channels is not so large and this method has been shown to be quite efficient.

\bibliographystyle{apsrev}	
\bibliography{refs}		

\end{document}